# Determination and evaluation of the critical liquid nitrogen for superconducting levitator based on a novel temperature-weight coupling measurement device

**Peng Pang[1], Jun Zheng[2], Chenling Xian[2]**

[1] School of Electrical Engineering, Southwest Jiaotong University, Chengdu 610031 People's Republic of China

[2] State Key Laboratory of Rail Transit Vehicle System, Southwest Jiaotong University, Chengdu 610031 People's Republic of China

Email: jzheng@swjtu.edu.cn

## Abstract

Liquid nitrogen ($LN_2$) is the only cooling medium for the high-temperature superconducting (HTS) bulks in the superconducting levitator, which is the heart of the maglev train, to reach working state. The detection and determination of the critical $LN_2$ content are crucial for reliable operation of the HTS maglev train. However, the related intelligent detection model and technology is lack in the combination filed of the cryogenic environment and maglev application, and there is no existing method to detect the $LN_2$ content in superconducting levitator. This paper proposes to employ multisensor fusion framework to fuse and enhance the accuracy of critical $LN_2$ content testing. Four temperature sensors were deployed inside superconducting levitator to measure the temperature change during the $LN_2$ content changing from 100 % to 0. It was first obtained that the critical $LN_2$ content in the superconducting levitator is 4%. To accurately monitor the critical $LN_2$ content in the superconducting levitator, a matrix-weighted information fusion Kalman filter algorithm was used. Compared with the previous single sensor method, the testing accuracy of the multisensor fusion method can be improved by 5.6%. The work can provide a preliminary research foundation for the online monitoring and fault diagnosis of HTS maglev train.

**Keywords:** High-temperature superconducting bulk, Liquid nitrogen content, Multisensor fusion, Online monitoring.

## I. Introduction

Magnetic levitation (maglev) trains are more suitable for high-speed operation and are also an important direction for the development of advanced rail transit [1]-[5]. Among them, the high-temperature superconducting (HTS) maglev technology is one important branch of maglev technologies due to its ability to achieve passive self-stable levitation and no magnetic resistance in the direction of running. The entire HTS maglev system mainly consists of two parts: HTS bulks and permanent magnet guideway (PMG) [6]-[8]. The low-temperature liquid nitrogen ($LN_2$: 77 K) is used to cool the HTS bulk below the critical temperature 93 K [9], allowing it to enter the superconducting state. The HTS bulk can achieve self-stable levitation exposed to an external magnetic field in the superconducting state. Once the temperature of the HTS bulk exceeds 93 K, it will lose its superconducting characteristics and the HTS maglev system will fail immediately [10][11], and the trains will hit the rail. These phenomena seriously affect the safety and punctuality of the maglev trains. Therefore, studying the minimum $LN_2$ content and corresponding detection methods of superconducting levitator is crucial for the safe operation of HTS maglev train.

The structure of the HTS maglev train and superconducting levitator is shown in Fig. 1. The HTS bulks are encapsulated at the bottom of the superconducting levitator [12][13]. The $LN_2$ container (low-temperature cavity) in the superconducting levitator uses $LN_2$ as a coolant to provide a cryogenic environment for the HTS bulk array by the cooling guide plate. During the operation of the maglev train, it is necessary to ensure that there is sufficient $LN_2$ in the superconducting levitator to cool the HTS bulks. However, the $LN_2$ within the low-temperature cavity undergoes continuous evaporation, primarily because of the uninterrupted connection between the $LN_2$ injection port and the atmosphere. Consequently, the $LN_2$ content steadily diminishes over time. Additionally, during the operation of HTS maglev train, the increase in speed and the AC losses will also exacerbate the volatilization of $LN_2$ [14]. The lack of $LN_2$ can lead to insufficient cooling capacity for the HTS bulk array, resulting in a temperature rise of the HTS bulks.

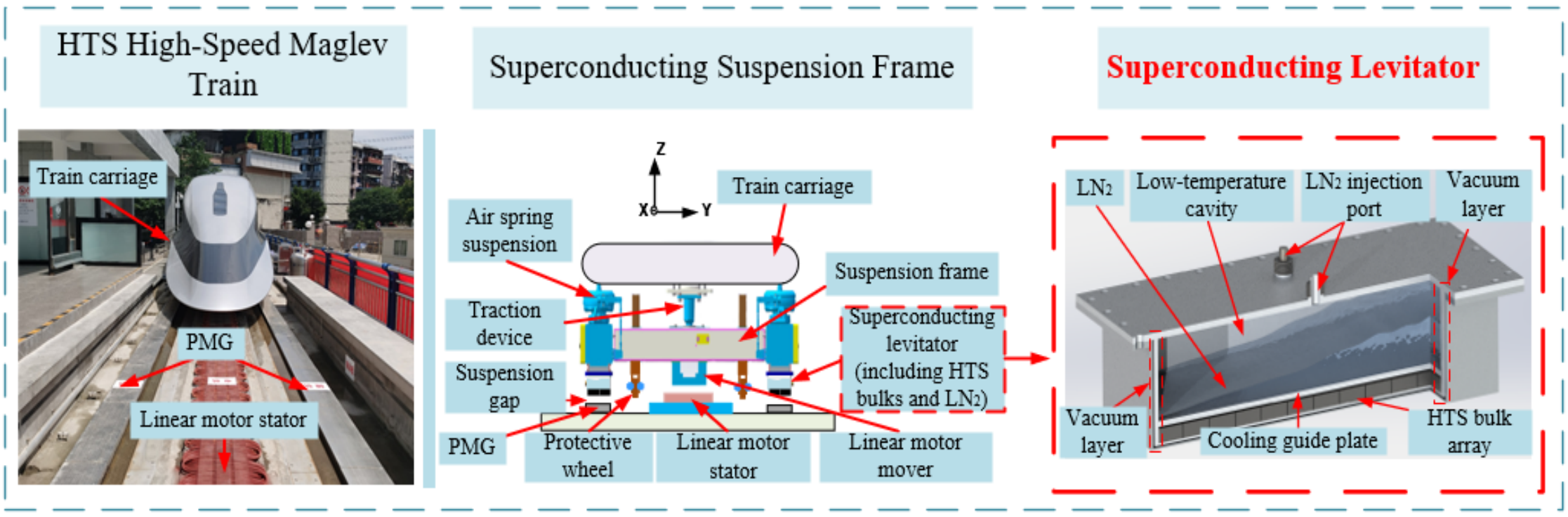

**Fig. 1.** The structure of the HTS maglev train and superconducting levitator.

Therefore, the adequate $LN_2$ content in the superconducting levitator is crucial for maintaining the superconducting state of the HTS bulks. In addition to the HTS maglev train, the $LN_2$ content is also very important in other HTS systems, like HTS magnets [15]-[19] and HTS bearings [20]-[22]. Junil Kim et al. [15] analyzed the effect of different $LN_2$ contents on the dielectric properties of HTS magnets. S Bang et al. [16] have published recommendations and evaluation work on the selection criteria of $LN_2$ level in the low-temperature field of superconducting coil magnets. For HTS magnetic bearing, A J C Arsénio et al. [20] also demonstrated that insufficient $LN_2$ content can cause a faster temperature rise of the HTS bulk under the $LN_2$ conduction cooling method. Hence, it is imperative to employ precise methods for testing the $LN_2$ content in HTS systems that rely on $LN_2$ cooling.

The superconducting levitator is a closed low-temperature container, it is not possible to observe the $LN_2$ content directly. The working environment inside the superconducting levitator that spans a large temperature range (room temperature 298 K to deep cooling 77 K) and vacuum environment are also not encountered in other liquid-level detection situations. Researchers have proposed specific testing methods for determining the $LN_2$ content in the superconducting levitator [23][24].

P Wen et al. [23] used a temperature sensor to preliminarily calculate the $LN_2$ content under continuous evaporation using the weight change during the process of $LN_2$ content change in the cryostat. However, due to the harsh operating conditions of the superconducting levitator, relying on the information provided by a single sensor makes it difficult to meet the accuracy requirements of state estimation. At the same time, due to random factors such as the location of sensors, differences in sensor quality, and working environment, the measurement results of each sensor cannot fully reflect the true result. S Zhang et al. [24] used 8 temperature sensors to measure the $LN_2$ content and obtained the temperature values of each temperature sensor as the $LN_2$ content continued to change, but threshold judgment cannot accurately measure the true $LN_2$ content. Although they [23][24] have proposed corresponding methods for $LN_2$ level detection, the collected data is not from a real superconducting levitator. The existing temperature sensor installation and testing methods cannot effectively evaluate the $LN_2$ content inside the superconducting levitator, and new testing method need to be explored.

The multisensor information fusion method has smaller errors and higher accuracy in status assessment, therefore, the multisensor information fusion method provides a research method for measuring the $LN_2$ content in the superconducting levitator since it also plays an important role in other systems [25]-[27]. M Fu et al. [28] utilized a multisensor information fusion method to extract magnetic leakage information, which can accurately characterize the true distribution of the geomagnetic field. S Wan et al. [29] utilized multisensor information fusion network algorithms to accurately identify bearing faults by processing different physical signals of bearings. Y Tang et al. [30] proposed a motor fault diagnosis method based on visual features driven by multiple sensors, and their experiments had shown that this method has high efficiency.

To get and achieve accurate measurement of the critical $LN_2$ content in the superconducting levitator, this paper establishes a comprehensive testing platform for measuring both weight and temperature inside the superconducting levitator, and proposes a testing method based on multisensor information fusion for the $LN_2$ content. The feasibility and accuracy of measuring the critical $LN_2$ content by positioning temperature sensors within the vacuum layer have been validated for the first time. The change of temperature sensors is recorded as the $LN_2$ content changes from 100% to 0. Then, optimal Kalman filtering was used to construct state and observation equations for each temperature sensor. Finally, $LN_2$ content testing method based on three temperature sensors was constructed using the matrix weighting method, and compared with the results of single temperature sensor. More specifically, the contributions of this paper are concluded as follows:

1) The self-developed SCML-03 device [31] and the temperature rise measurement equipment are used to build a novel experimental device. This new device can simultaneously measure the weight and internal temperature rise of the superconducting levitator.

2) This paper innovatively proposes a method for testing $LN_2$ content by placing temperature sensors in a real superconducting levitator, and obtains changes in the $LN_2$ content inside the superconducting levitator with time. In addition, the $LN_2$ content at which the HTS bulk begins to experience temperature rise is also obtained.

3) Comparison analysis was conducted between single sensor and information fusion testing. Finally, from the perspective of cost and system complexity, this paper proposes targeted suggestions for the detection of $LN_2$ content in the superconducting levitator.

The structure of this paper is arranged as follows: The testing platform and filtering algorithms are introduced in section II. The experimental result and data processing are described in section III. Finally, the main conclusions of this paper are summarized in section IV.

# II. Methodology

Unlike previous $LN_2$ content testing platforms, this study selected the full-size superconducting levitator equipped with HTS maglev train as the research object as shown in Fig. 2. At the same time, by combining the self-developed SCML-03 and the dynamic temperature rise testing device, it is possible to accurately measure the changes of the temperature sensors inside the superconducting levitator as the $LN_2$ content changes.

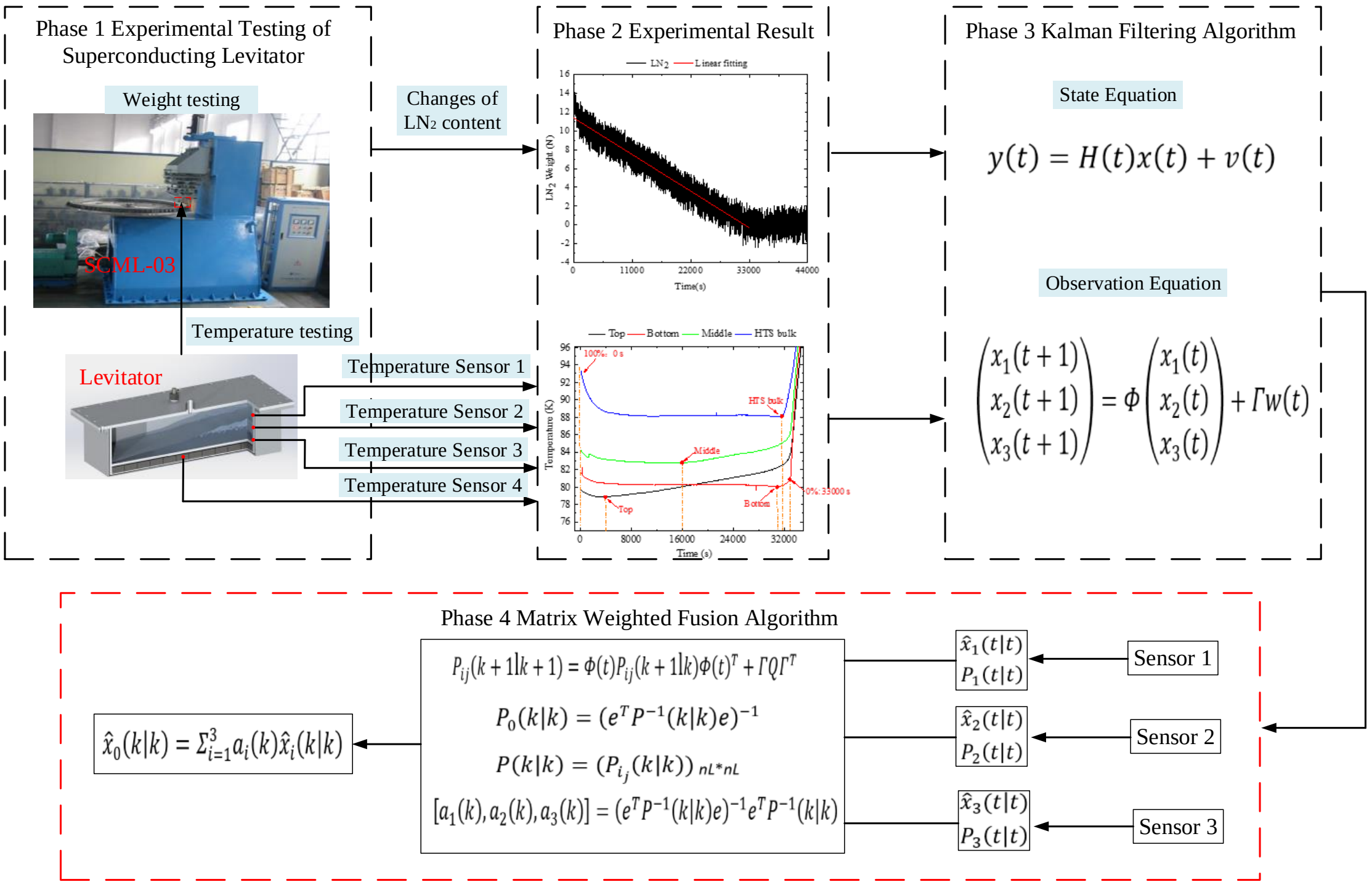

**Fig. 2** Testing framework for $LN_2$ content in superconducting levitator based on multisensor information fusion.

## A. The Equipment and process of the experiment

The SCML-03 device [31] was originally designed to record the changes in levitation and guidance force of the HTS maglev system exposed to dynamic magnetic field fluctuations. The HTS bulks are fixed on the upper part of the circular PMG and it can simulate the actual operating conditions of the HTS maglev trains when the circular PMG rotates. In this paper, the superconducting levitator was installed on the SCML-03 device with the circular PMG kept stationary, and the change in the weight of the superconducting levitator during the evaporation of the $LN_2$ was recorded. The installation of the superconducting levitator is shown in Fig. 3. The change in weight of the superconducting levitator is the change of the $LN_2$ content.

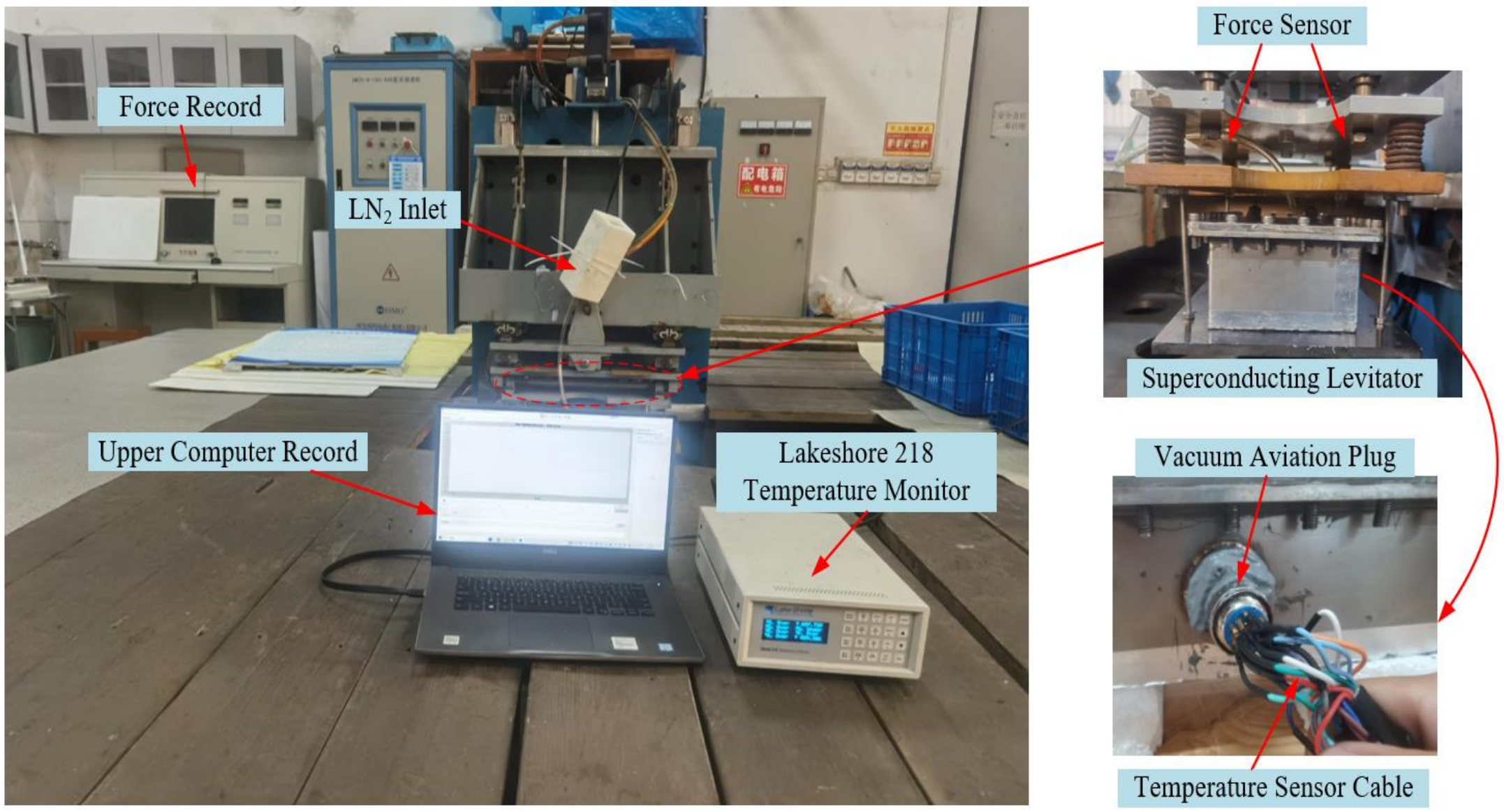

**Fig. 3.** The testing process of $LN_2$ content in the superconducting levitator on SCML-03.

The experimental process is as follows: The aviation plug is installed on the outer wall of the superconducting levitator, and the temperature sensors are fixed inside the superconducting levitator. Then leading out the wires through the aviation plug and put the superconducting levitator on SCML-03. $LN_2$ is added to the inside of the superconducting levitator through the $LN_2$ inlet. Finally, turning on the force record of the SCML-03 and temperature collection device to record the weight and temperature changes during the process of $LN_2$ content changes in the superconducting levitator. Initially, by tracking the variations in weight of the superconducting levitator during $LN_2$ evaporation, a state equation is formulated to precisely describe the correlation between its state and weight alterations. The observation equation is devised based on data from three temperature sensors affixed to the outer wall of the low-temperature cavity, while the temperature sensor positioned at the bottom of the HTS bulk is employed to ascertain the critical $LN_2$ content. When a temperature increase is detected by the bottom temperature sensor of the HTS bulk, it indicates that the $LN_2$ content has reached the critical level. Subsequently, a matrix fusion algorithm is deployed for monitoring the $LN_2$ content of the superconducting levitator. Thus, the process involves constructing observation equations using three temperature sensors and subsequently employing a matrix fusion algorithm for monitoring the $LN_2$ content.

## B. Optimal Kalman filtering

Kalman filter is mainly used for predicting and updating the status of the system, which offers several advantages over other filtering methods [32][33]. Firstly, it combines both prediction and correction steps recursively, allowing for real-time estimation of system states even in the presence of noise or uncertainty. Secondly, it optimally incorporates information from both current sensor measurements and prior knowledge about the system's dynamics, leading to more accurate and reliable state estimates. Moreover, it provides a rigorous framework for handling non-linear systems. Therefore, this paper uses the optimal Kalman filter to handle the testing results of the superconducting levitator, and the observation values are used to calculate and estimate the optimal state of the system. According to the Kalman filtering principle, the equations for a stochastic linear

discrete system are:

$$\boldsymbol{X}(t+1) = \Phi \boldsymbol{X}(t) + \boldsymbol{\Gamma W}(t) \quad (1)$$

$$\boldsymbol{Y}(t) = H\boldsymbol{X}(t) + \boldsymbol{V}(t) \quad (2)$$

where $\boldsymbol{X}(t) \in R^n$ is the state of the system at time t, $\boldsymbol{Y}(t) \in R^m$ is the observed state signal, $\boldsymbol{W}(t) \in R^r$ is the input white noise, and $\boldsymbol{V}(t) \in R^n$ is the observed noise. Equation (1) represents the equation of state, (2) represents the observation equation, $\Phi$ is the state transition matrix, and $H$ is the observation matrix. The following assumptions need to be made before conducting Kalman filtering calculations as following:

1) $\boldsymbol{W}(t)$ and $\boldsymbol{V}(t)$ are uncorrelated white noise with a mean of 0 and variances of $\boldsymbol{Q}$ and $\boldsymbol{R}$, respectively. E$\boldsymbol{W}(t)$=0, E$\boldsymbol{V}(t)$=0, E[$\boldsymbol{W}(t)\boldsymbol{W}^T(t)$]=$\boldsymbol{Q}\delta_{ij}$, E[$\boldsymbol{V}(t)\boldsymbol{V}^T(t)$]=$\boldsymbol{R}\delta_{ij}$, E[$\boldsymbol{W}(t)\boldsymbol{V}^T(t)$]=0, E represents the expected calculation. ∀i, j, $\delta_{ii}$=1, $\delta_{ij}$=0.
2) $\boldsymbol{X}(0)$ is not related to $\boldsymbol{W}(t)$ and $\boldsymbol{V}(t)$, E[$\boldsymbol{X}(0)$]=$\boldsymbol{\mu}_0$, E[($\boldsymbol{X}(0)$-$\boldsymbol{\mu}_0$) ($\boldsymbol{X}(0)$-$\boldsymbol{\mu}_0$)$^T$]=$\boldsymbol{P}_0$.

The prediction process for the optimal Kalman filter of the No. i sensor is as follows:

State recursive predictor:

$$\boldsymbol{X}_i(t+1\text{l}t) = \boldsymbol{\psi_P}(t)\boldsymbol{X}_i(t\text{l}t-1) + \boldsymbol{k}_{\boldsymbol{p}_i}(t)\boldsymbol{Y}(t) \quad (3)$$

Gain:

$$\boldsymbol{k_{fi}}(t) = \boldsymbol{P_{ii}}(t\text{l}t-1)\boldsymbol{H^T}(t)[\boldsymbol{H}(t)\boldsymbol{P_{ii}}(t\text{l}t-1)\boldsymbol{H}^T(t) + \boldsymbol{Q_v}(t)]^{-1} \quad (4)$$

$$\boldsymbol{k_{pi}}(t) = \Phi \boldsymbol{k_{fi}}(t) \quad (5)$$

Error equation matrix:

$$\boldsymbol{P_{ii}}(\text{t}+1\text{lt}) = \boldsymbol{\psi_P}(t)\boldsymbol{P_{ii}}(tlt-1)\boldsymbol{\psi_P}(t)^T + \boldsymbol{k_{pi}}(t)\boldsymbol{Q_v}(t)\boldsymbol{k_{pi}}(t)^T + \boldsymbol{\Gamma}(t)\boldsymbol{Q_v}(t)\boldsymbol{\Gamma}(t)^T \quad (6)$$

$\boldsymbol{k_{fi}}(t)$: the Kalman gain;

$\boldsymbol{P_{ii}}(t+1\text{lt})$: the variance matrix;

$\boldsymbol{Q_v}(t)$: the covariance matrix;

$\boldsymbol{k_{pi}}(t)$ : the gain matrix;

$\boldsymbol{\psi_P}(t)$: the error matrix;

*T:* the transposed symbol.

### C. Matrix Weighted Fusion Algorithm

Although the $LN_2$ content measured by each sensor can obtain a relatively stable and accurate value after Kalman filtering, a single sensor may malfunction or be interfered with during $LN_2$ content measurement, resulting in a significant difference between the obtained value and the true value. Furthermore, the $LN_2$ content measured by a single temperature sensor may encounter time delays during transmission and exchange in the actual $LN_2$ content process. Therefore, to reduce potential deviations during information processing and improve the accuracy, it is necessary to use multiple sensors to measure the $LN_2$ content. The predicted values of each sensor after Kalman filtering can be obtained through (1)-(6). The cross-covariance of the estimation error between any two sensors is:

$$\boldsymbol{P_{ij}}(k+1\text{l}k+1) = \Phi \boldsymbol{P_{ij}}(k+1\text{l}k)\Phi^T + \boldsymbol{\Gamma Q \Gamma^T} \quad (7)$$

The block matrix of $\boldsymbol{P_{ij}}(k+1\text{l}k+1)$ is:

$$\boldsymbol{P}(k|k) = (\boldsymbol{P}_{i_j}(k|k)) \text{ nL*nL} \quad (8)$$

The optimal weighting matrix is:

$$[a_1(k), \dots, a_L(k)] = (\mathrm{e}^{\boldsymbol{T}}\boldsymbol{P}^{-\mathbf{1}}(k|k)\mathrm{e})^{-1}\mathrm{e}^{\boldsymbol{T}}\boldsymbol{P}^{-\mathbf{1}}(k|k) \tag{9}$$

The estimated value of matrix weighted fusion is:

$$\boldsymbol{X}_{\mathbf{0}}(k|k) = \Sigma_{\mathrm{i}=1}^{l} a_i(k)\boldsymbol{X}_{\boldsymbol{i}}(k|k) \tag{10}$$

The optimal fusion error matrix is:

$$\boldsymbol{P}_{\mathbf{0}}(k|k) = (\mathrm{e}^{\boldsymbol{T}}\boldsymbol{P}^{-\mathbf{1}}(k|k)\mathrm{e})^{-1} \tag{11}$$

In the experiment, the initial values of $LN_2$ content and the initial changes in $LN_2$ content are allocated to the initial state, and an estimated initial covariance is provided. Subsequently, these values are inputted into the optimal Kalman matrix fusion filter to derive stable and accurate real $LN_2$ content.

# III. EXPERIMENTAL RESULTS

The testing result of the $LN_2$ content changes in the superconducting levitator is shown in Fig. 4. The $LN_2$ in the superconducting levitator remained for about 33000 s during the process from 100% to 0. Due to the large volatility caused by the volatilization of $LN_2$ during the testing process, linear fitting was performed on the original data, as shown by the red line in Fig. 4. In addition, after the $LN_2$ in the superconducting levitator completely evaporated, the levitator was filled with nitrogen gas, and the $LN_2$ content does not completely become 0.

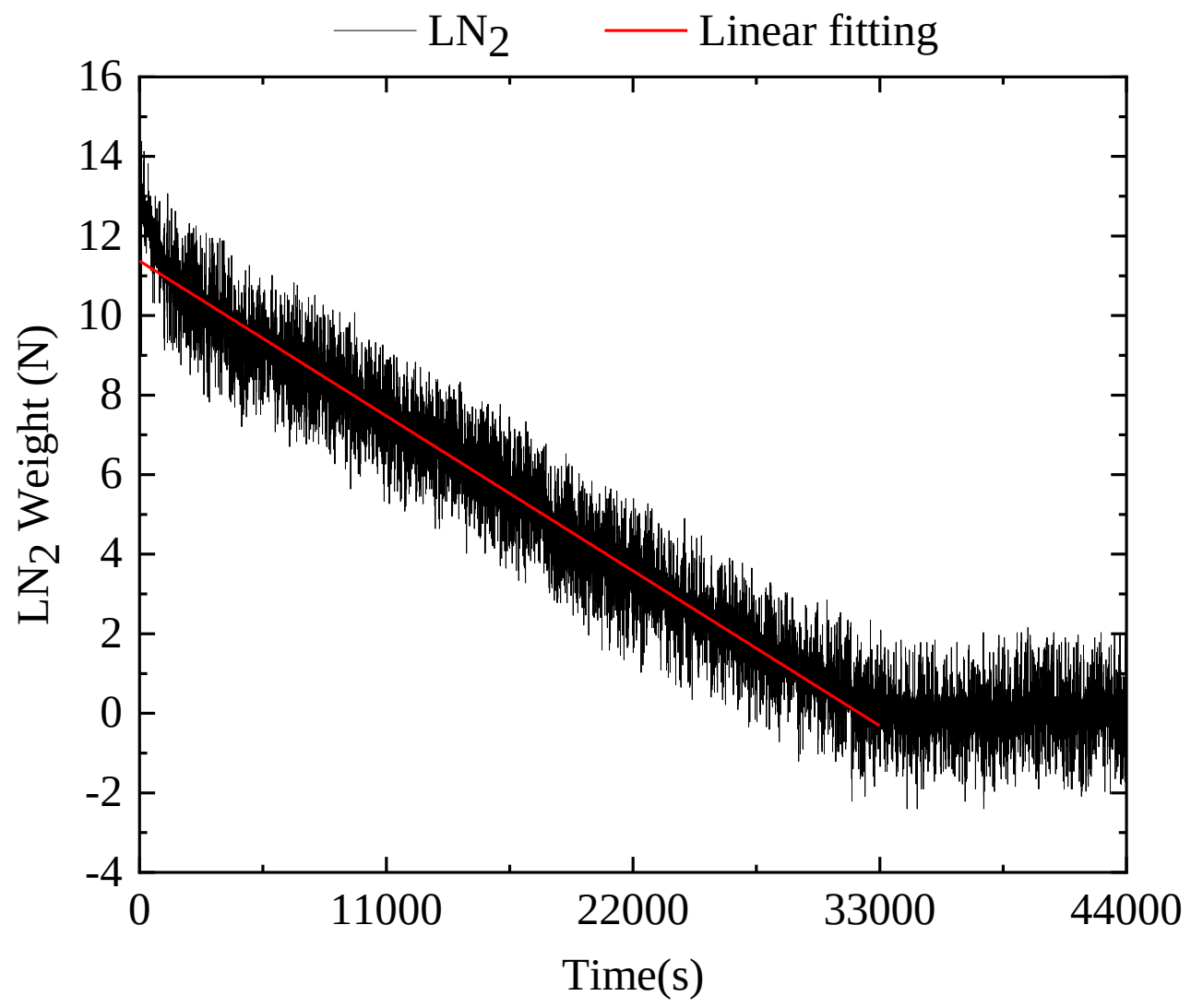

**Fig. 4.** The change of $LN_2$ content in superconducting levitator.

Based on the linear fitting results, the relationship between the $LN_2$ content in the superconducting levitator and time is obtained as follows:

$$Y = 1 - 0.0000303t \tag{12}$$

The testing results of four temperature sensors in the superconducting levitator are shown in Fig. 5. The arrow in Fig. 5 represents the moment when the corresponding position begins to experience temperature rise. The values of the temperature sensors slowly decrease because they are attached to the inner wall of the low-temperature cavity in the superconducting levitator after filling the $LN_2$. Due to the different thicknesses of the black glue between the temperature sensors and the inner wall, the final stability values of the three temperature sensors on the inner wall are different.

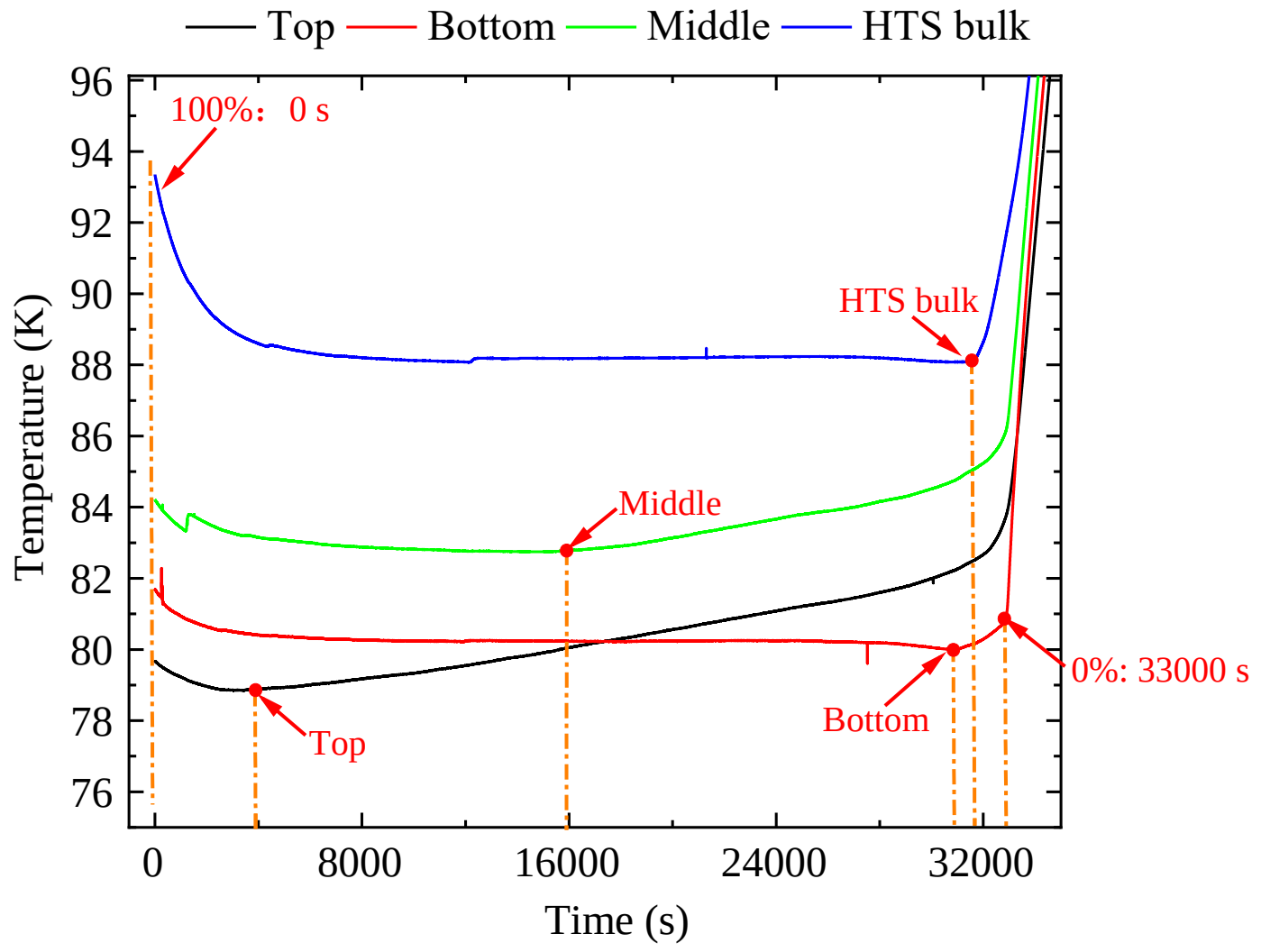

**Fig. 5.** Changes in each temperature sensor over time in the superconducting levitator.

The temperature of middle sensor is the highest, followed by the lower value, and the upper temperature sensor has the lowest value. Since the HTS bulk is at the lowest part of the entire superconducting levitator, the temperature value of the lower surface of the HTS bulk is the highest, which is 88 K. As the $LN_2$ content continues to decrease, the value of the top temperature sensor begins to rise at around 4000 s, and then, the value of the middle temperature sensor begins to rise at 16000 s, indicating that the $LN_2$ content in the superconducting levitator is less than half. As the $LN_2$ content continues to decrease, the value of the bottom temperature sensor begins to rise, and soon after, the temperature of the HTS bulk begins to rise. Finally, as the $LN_2$ evaporated completely, the temperature rise rate of the three temperature sensors increased sharply at 33000 s, proving that there was no $LN_2$ in the superconducting levitator.

Based on the temperature rise of the HTS bulk at the bottom of the superconducting levitator at 32000 s, when the $LN_2$ content is approximately 4%. To obtain the relationship between temperature sensors and $LN_2$ content, the data of each temperature sensor needs to be processed. Due to the approximate linearity of each temperature sensor during the temperature rise stage as shown by the red line in Fig. 6. For subsequent analysis, each temperature sensor is linearized separately during the rise stage by the blue line in Fig. 6, and the linearization results of each temperature sensor are:

$$\text{Top:}\ T_{top} = 82.06 + 0.00012t \tag{13}$$

$$\text{Middle:}\ T_{middle} = 84.78 + 0.00015t \tag{14}$$

$$\text{Bottom:}\ T_{bottom} = 79.97 + 0.00038t \tag{15}$$

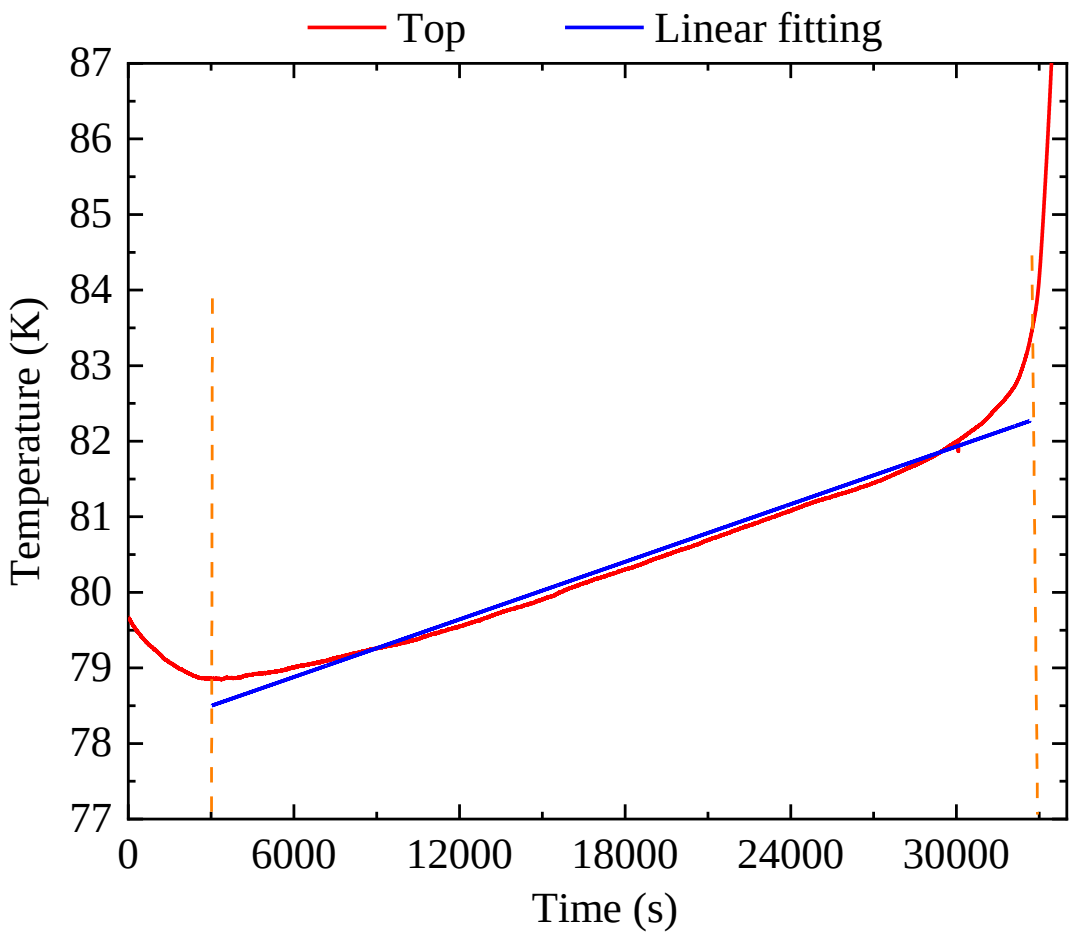

(a) Top temperature sensor

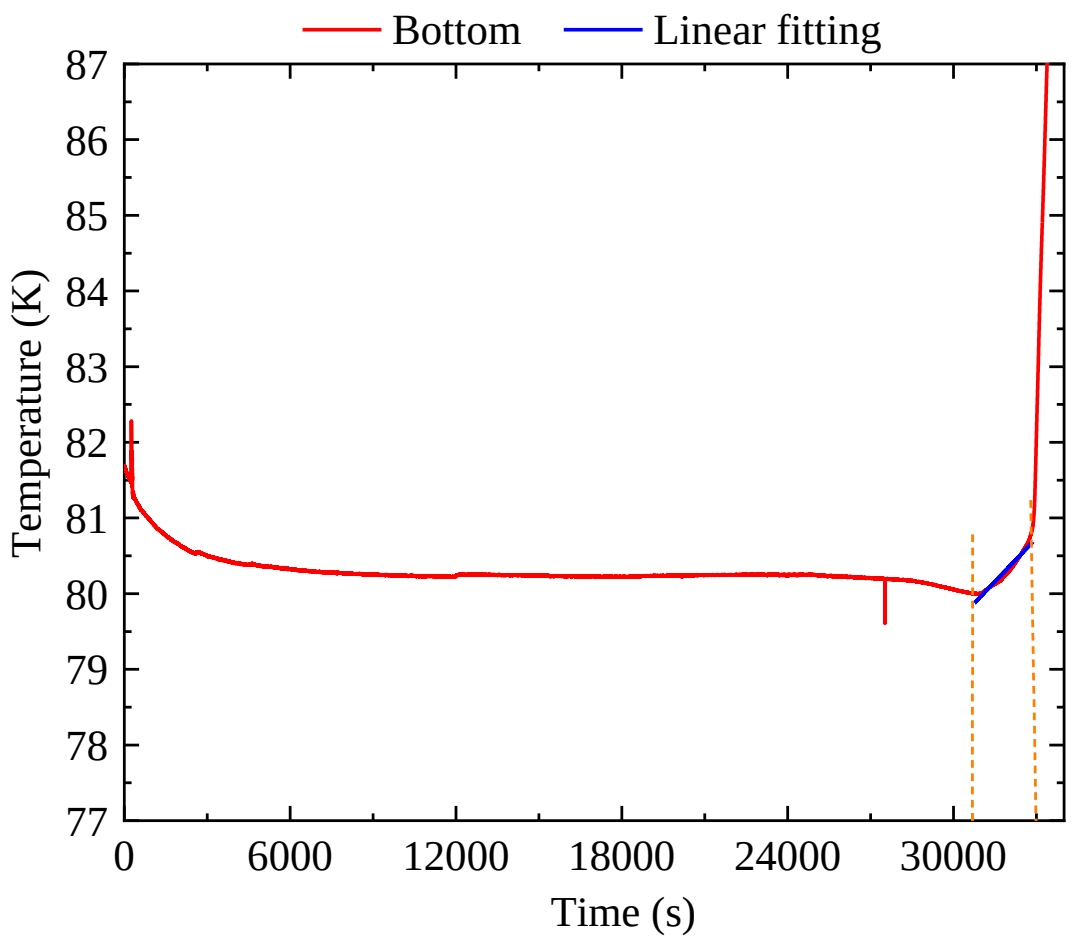

(b) Bottom temperature sensor

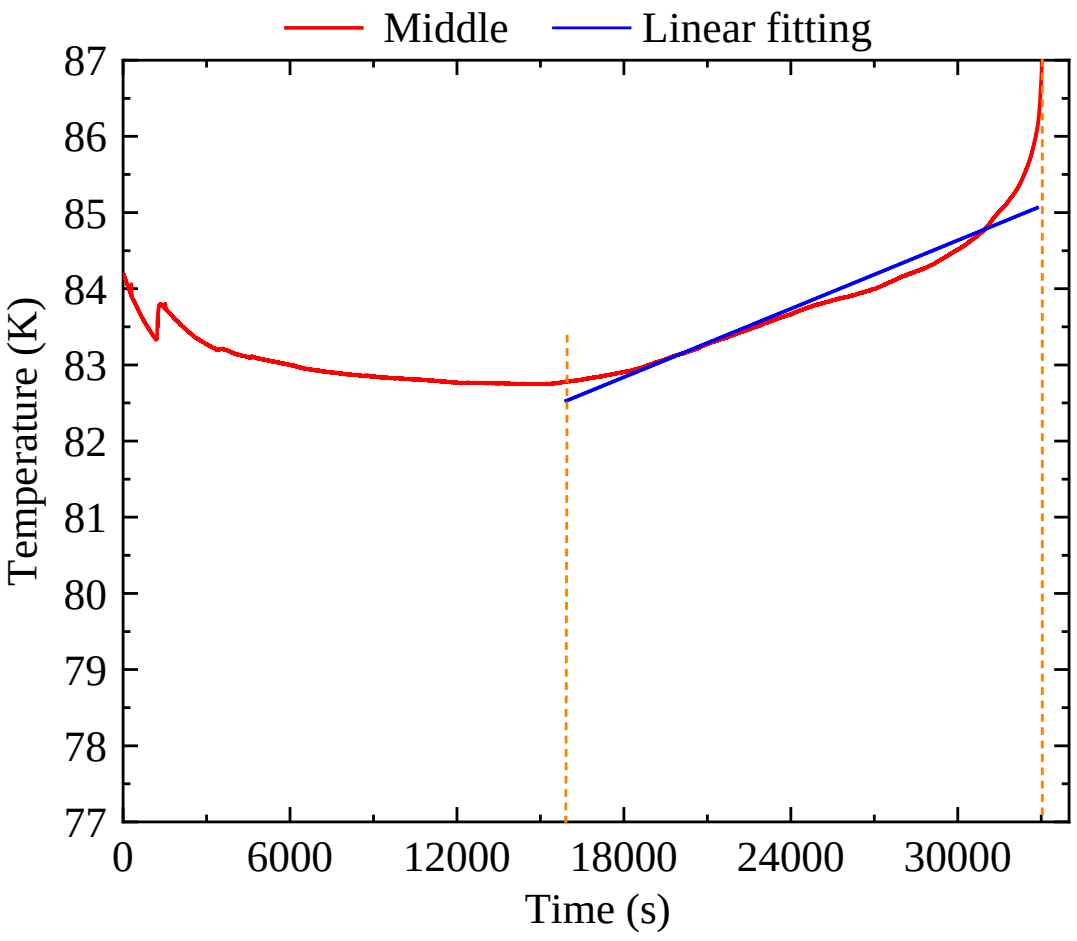

(c) Middle temperature sensor

**Fig. 6.** The linearization results of three temperature sensors.

From the results in Fig. 5, the temperature of the HTS bulk begins to rise when the $LN_2$ content is 4%, and the HTS bulk tends to quench. This paper is based on the numerical changes of each temperature sensor at 4% $LN_2$ content, and uses equations (1) - (10) to construct a critical $LN_2$ content testing method for superconducting levitator. According to equations (12) - (15), the initial conditions can be obtained:

$$\Phi = \begin{bmatrix} 1 & 0 & 0 \\ 0 & 1 & 0 \\ 0 & 0 & 1 \end{bmatrix}$$

$$\Gamma = [1,1,1]'$$

$$H = \begin{bmatrix} -0.048 & 0 & 0 \\ 0 & -0.06 & 0 \\ 0 & 0 & -0.048 \end{bmatrix}$$

Substitute the initial conditions into equations (1) - (10) to obtain the $LN_2$ content test results of a single temperature sensor, as shown in Fig. 7.

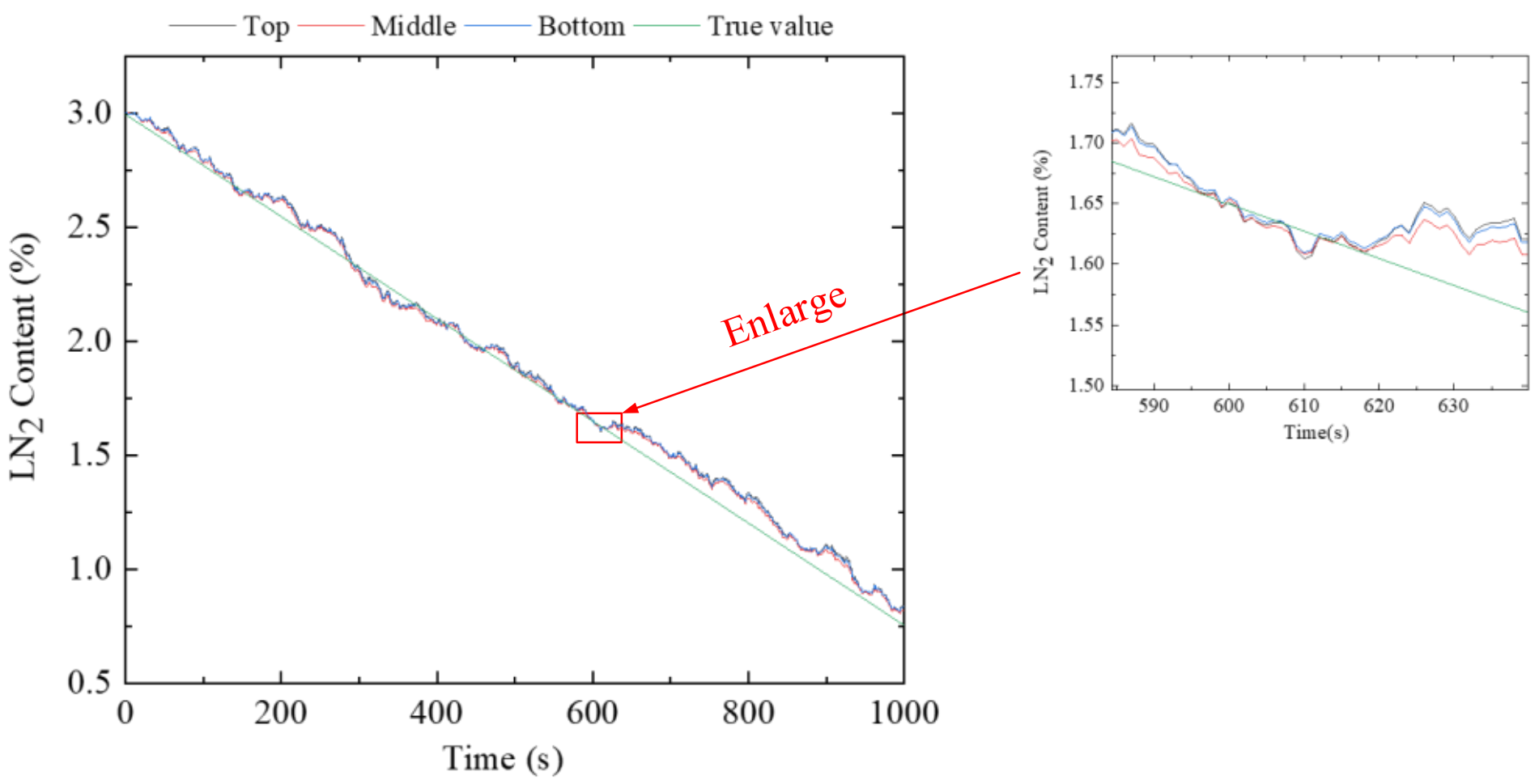

**Fig. 7.** The diagram of single sensor Kalman filtering results.

From the filtering results in Fig. 7, the filtering values of each temperature sensor are near the true values, indicating the effectiveness of Kalman filtering in testing $LN_2$ content in the superconducting levitator. To improve the accuracy of measurement and anti-interference ability, the filtering results obtained by using the multi-sensor information fusion method are compared with the filtering results of single sensor, as shown in Fig. 8.

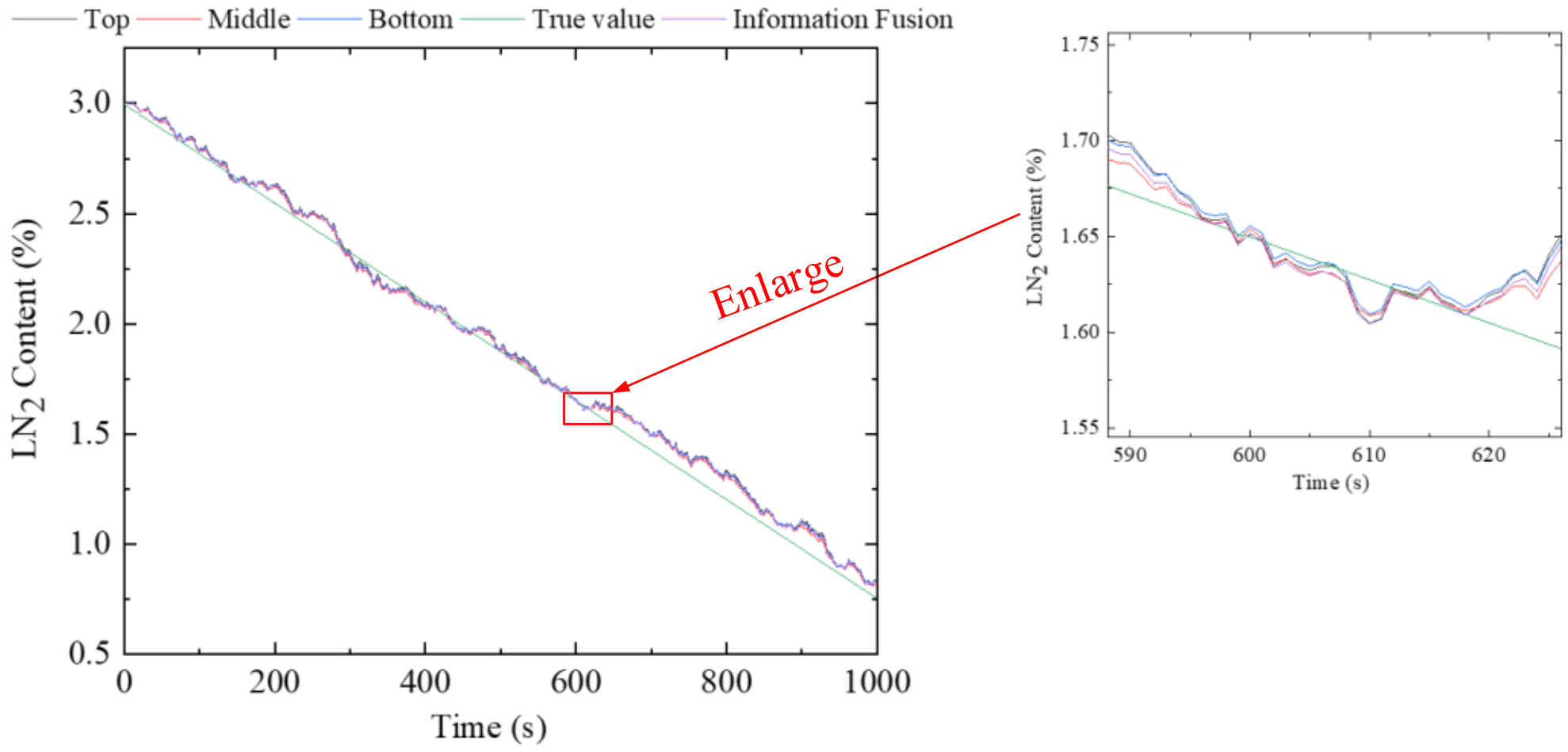

**Fig. 8.** Comparison of multisensor fusion and single sensor filtering results.

From the filtering results in Fig. 8, the results obtained by the multisensor fusion algorithm are closer to the real results. The error calculation formula is:

$$Error = L_{real} - L_{test} \tag{16}$$

where $L_{real}$ is the real value of the $LN_2$ content, and $L_{test}$ is the filter value. The errors of single sensor and multi-sensor fusion are shown in Fig. 9, and table I shows the average error comparison. The filtering values fluctuate continuously with time as shown in Fig. 9. Moreover, average error in descending order is: multi-sensor fusion, middle sensor, bottom sensor, and top sensor. The accuracy of the middle temperature sensor is the highest as shown in table I, which is 14.5% and 10.8% higher than that of the top and bottom temperature sensors. So, the middle temperature sensor has the highest testing accuracy in single sensor testing, followed by the bottom sensor, and the top sensor has the lowest testing accuracy. The testing accuracy of multiple sensors has been improved by 5.6% when compared to the middle sensor. The result clearly indicates that the multisensor information fusion method produces more accurate results.

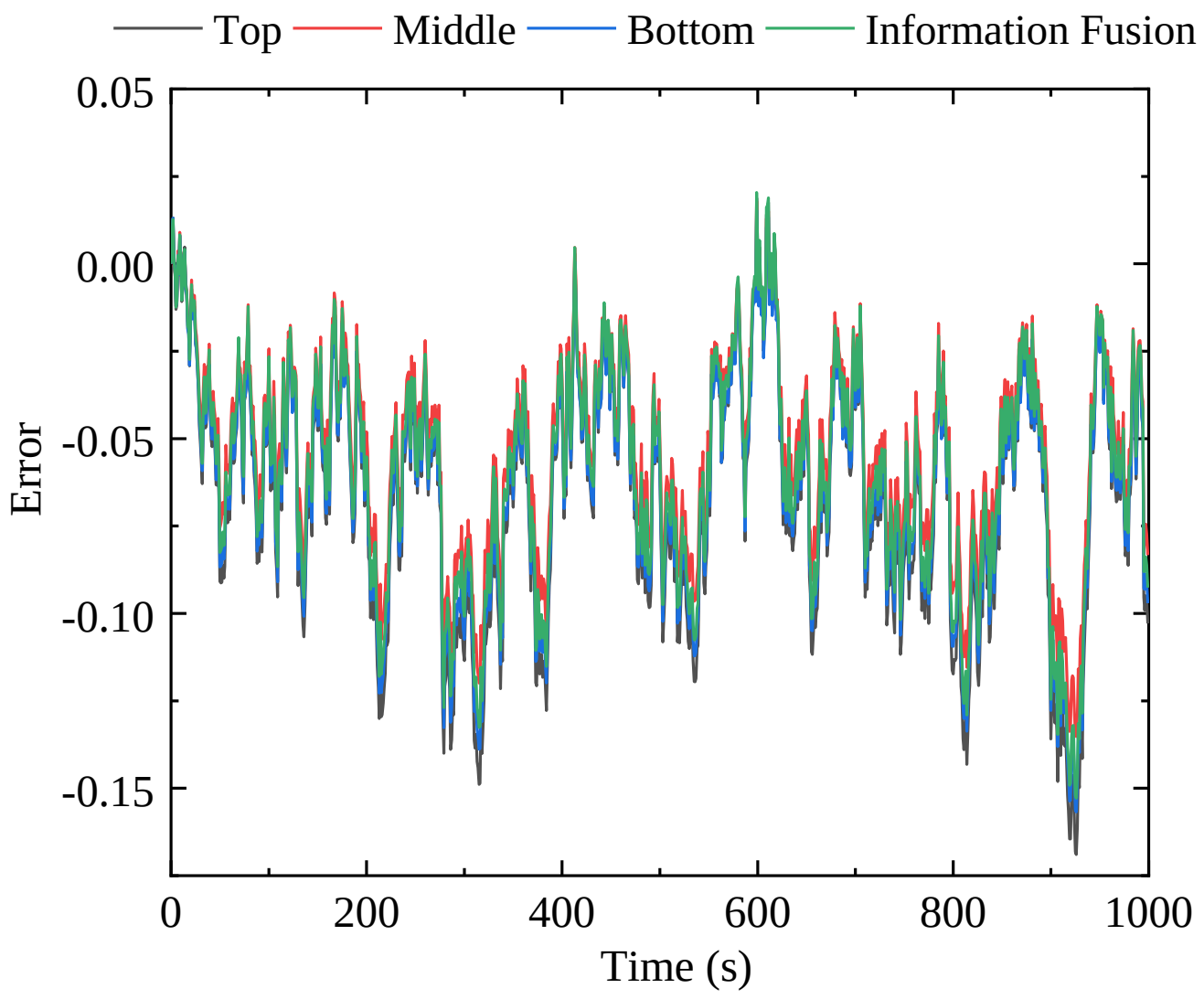

**Fig. 9.** The comparation of error analysis.

Table I AVERAGE ERROR RESULTS

| Different filtering methods | Average Error |
|---|---|
| Top temperature sensor | 0.06603 |
| Middle temperature sensor | 0.05645 |
| Bottom temperature sensor | 0.06325 |
| Multisensor information fusion | 0.05328 |

# IV. Conclusion

In this paper, a method for measuring the critical $LN_2$ content in the superconducting levitator based on multisensor information is proposed. Firstly, this paper used the method of placing temperature sensors in the real superconducting levitator to test the $LN_2$ content. Then, the relationship between the $LN_2$ content in the superconducting levitator and the temperature sensors on the outer wall of the low-temperature cavity was explored, and a state equation was constructed. Finally, a multisensor information fusion algorithm was introduced for testing the critical $LN_2$ content, proving that the multisensor method has higher accuracy for $LN_2$ content testing. The main conclusions of this paper are as follows:

(1) Based on the temperature sensors, it is evident that the temperature at the bottom of the HTS bulk inside the superconducting levitator is 88 K, which is significantly different from the $LN_2$ temperature of 77 K. This is because the current superconducting levitator uses stainless steel as the cooling plate, which has a relatively low heat transfer rate.

(2) For the current structure of the superconducting levitator, the accuracy of $LN_2$ content testing using a single temperature sensor is relatively low. The use of three temperature sensors combined with information fusion algorithms can meet more accurate $LN_2$ content testing requirements.

(3) Three temperature sensors were placed in the superconducting levitator for $LN_2$ content testing, which occupied a limited inner space. Therefore, to reduce cost and detection complexity in the future, a temperature sensor can be placed in the middle position for detection. Although this approach may not achieve the same high accuracy as multisensor, it still provides relatively reliable results.

In the next step of research, the critical $LN_2$ content testing method for superconducting levitator based on multisensor information fusion can be considered in the actual operating state of HTS maglev train, and explore the real-time testing accuracy under different operating conditions. Meanwhile, there are multiple HTS bulks at the bottom of the superconducting levitator. In the future, temperature sensors need to be placed at different positions of the HTS bulks to explore the optimal critical $LN_2$ content of the superconducting levitator.

# CRediT authorship contribution statement

**Peng Pang**: Conceptualization, Data curation, Methodology, Software, Writing – original draft. **Chenling Xian**: Investigation, Data curation. **Jun Zheng**: Conceptualization, Methodology, Resources, Supervision, Writing – review & editing, Funding acquisition.

# Declaration of Competing Interest

The authors declare that they have no known competing financial interests or personal relationships that could have appeared to influence the work reported in this paper.

# Acknowledgements

This work was partially supported by the National Natural Science Foundation of China (52375132, 52077178), the Sichuan Provincial Science and Technology Program (2022JDTD0011) and the Fundamental Research Funds for the Central Universities (2682023ZTPY040).